\documentclass[aps,prl,showpacs,twocolumn,floats]{revtex4}
\usepackage{graphicx}
\usepackage{dcolumn}
\usepackage{bm}
\begin{document}

\title{Single magnetic molecule between conducting leads: Effect of mechanical rotations}
\author{Reem Jaafar, Eugene M. Chudnovsky and Dmitry A. Garanin}
\affiliation{Physics Department, Lehman College, City University
of New York \\ 250 Bedford Park Boulevard West, Bronx, New York
10468-1589, USA}
\date{\today}
\begin{abstract}
We study spin-rotation effects in a magnetic molecule bridged
between two conducting leads. Dynamics of the total angular
momentum couples spin tunneling to the mechanical rotations.
Landau-Zener spin transition produced by the time-dependent
magnetic field generates a unique pattern of mechanical
oscillations that can be detected by measuring the electronic
tunneling current through the molecule.
\end{abstract}
 \maketitle

In the last years a significant experimental effort has been made
to measure electronic current through a single magnetic molecule
bridged between conducting leads. These studies have been driven
by possible applications of single-molecule magnets in
spintronics, as well as by the hope to use magnetic molecules as
qubits \cite{Wernsdorfer}. Heersche et al. observed striking
voltage dependence of the current through a single-molecule magnet
Mn$_{12}$, with a complete current suppression and excitations of
negative differential conductance on the energy scale of the
anisotropy barrier \cite{Heersche}. Jo et al. measured the
magnetic field dependence of the electron tunneling spectrum in a
transistor incorporating a Mn$_{12}$ molecule \cite{Jo}. Henderson
et al. observed a Coulomb blockade effect by measuring conductance
through a single-molecule magnet Mn$_{12}$ \cite{Henderson}. Voss
et al. conducted experiments and developed theoretical model for
the dependence of the tunneling current on the orientation of the
Mn$_{12}$ molecule \cite{Voss}.

Various aspects of the electronic transport through magnetic
molecules have been investigated theoretically. The effect of the
exchange coupling between spins of conducting electrons and the
spin of the Mn$_{12}$ molecule has been studied by G.-H. Kim and
T.-S. Kim \cite{Kim-04}. Elste and Timm \cite{Elste} developed a
model for the Coulomb blockade in a transport through a single
magnetic molecule weakly coupled to magnetic and nonmagnetic
leads. They also studied the possibility of writing, storing, and
reading spin information in memory devices based upon
single-molecule magnets \cite{Timm-08}. Kondo effect in transport
through a single-molecule magnet strongly coupled to metallic
electrodes has been investigated by Romeike et al
\cite{Romeike-06}. The effect of spin Berry phase on electron
tunneling has been studied by Gonz\'{a}lez and Leuenberger
\cite{Leuenberger}. Misiorny et al. investigated magnetic
switching of the molecular spin by a spin-polarized current
\cite{Misiorny-07}, as well as tunneling magnetoresistance
\cite{Misiorny-09}. Cornaglia et al. have studied the effect on
the transport of a soft vibrating mode of the molecule
\cite{Cornaglia}. First principle DFT calculations of the electron
transport through a Mn$_{12}$ single-molecule magnet and of
spin-filtering effect have been performed by Barraza-Lopez et al.
\cite{Park}.
\begin{figure}[ht]
\begin{center}
\vspace{-1.2cm}
\includegraphics[width=67mm,angle=-90]{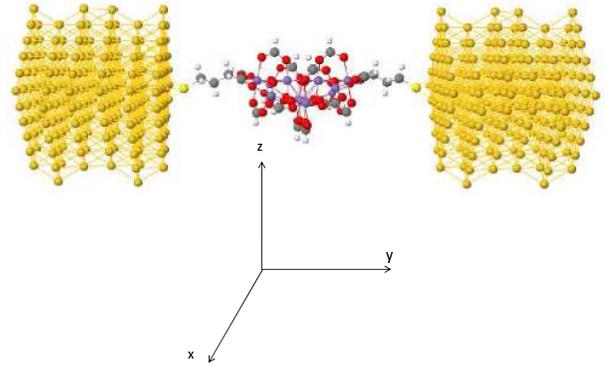}
\vspace{0cm} \caption{Mn$_{12}$ spin-$10$ magnetic molecule
bridged between metallic leads \cite{Park}.} \label{geometry}
\end{center}
\vspace{-0.3cm}
\end{figure}
There also has been some effort to compute Josephson current
through a magnetic molecule coupled to superconducting leads
\cite{JJ}.

Recently two of the authors have demonstrated \cite{CG-condmat}
that quantum states of a magnetic molecule that is free to rotate
are different from quantum states of a magnetic molecule whose
position is fixed in a crystal. This effect arises from the
conservation of the total angular momentum (spin + orbital).
Somewhat intermediate situation occurs when a molecule is bridged
between two electrodes. While such a molecule is not free to
rotate, its orientation can still vary due to the spin-rotation
coupling. In this Letter we study the Landau-Zener transition
between quantum states of the molecule caused by a field sweep. We
show that coupled dynamics of the molecular spin and the
mechanical rotation of the molecule leads to the modification of
the Landau-Zener dynamics.  It also results in a specific pattern
of mechanical oscillations of the molecule. Since electron
tunnelling through the molecule should strongly depend on its
orientation, this effect should be seen in the tunneling current.

We consider geometry depicted in Fig.\ \ref{geometry}. The
magnetic anisotropy axis of the molecule, $Z$, is perpendicular to
the direction of the transport current $Y$. The mechanical motion
of the molecule is restricted by its coupling to the leads. For
simplicity we consider only small torsional oscillations about the
$Z$-axis that we describe by the angle of rotation $\phi$. This is
justified by the fact that quantum tunneling of the molecular spin
between two opposite directions along the $Z$-axis changes only
the $Z$-component of the angular momentum, thus generating the
$Z$-component of the torque. In our approximation, small rotation
of the molecule due to this torque produces negligible deformation
of a rigid magnetic core of the molecule. On the contrary, the
tunneling current through the molecule must have exponential
dependence on $\phi$. We assume that electrons in the leads have
only marginal effect on the quantum spin states of the molecule.

The main part of the Hamiltonian of the molecule has three terms:
\begin{equation}\label{ham}
\hat{H}=\hat{H}_{rot} + \hat{H}_{Z}+ \hat{H}_{S}\,.
\end{equation}
Here
\begin{equation}
\hat{H}_{rot}=\frac{1}{2I_{z}}\left(\hbar^2
L_{z}^{2}+I_{z}^{2}\omega _{r}^{2}\phi ^{2}\right) \label{H-rot}
\end{equation}
describes mechanical rotations about the $Z$-axis. $I_z$ is the
corresponding moment of inertia and $\omega_r$ is the frequency of
free torsional oscillations of the molecule due its coupling to
the leads. Operator of the mechanical angular momentum, $L_{z}=-i
\partial /\partial \phi ,$ satisfies the commutation relation
\begin{equation}
\left[ \phi ,L_{z}\right] =i.  \label{RotComm}
\end{equation}
Interaction of the molecular spin ${\bf S}$ with the external
magnetic field, ${\bf B}$ applied along the $Z$-axis, is described
by the Zeeman term in Eq.\ (\ref{ham}):
\begin{equation}
\hat{H}_{Z}=-g\mu _{B}S_{z}B_{z}  \label{HZDef}\,,
\end{equation}
with $g$ being the gyromagnetic factor and $\mu_B$ being the Bohr
magneton. Operator $\hat{H}_{S}$ in Eq.\ (\ref{ham}) is given by
\cite{Dohm,CGS}
\begin{equation}
\hat{H}_S=\hat{R}\hat{H}_{A}\hat{R}^{-1}\,, \label{ham-S}
\end{equation}
where $\hat{H}_{A}$ is the crystal-field (magnetic anisotropy)
spin Hamiltonian and
\begin{equation}
\hat{R}=e^{-iS_{z}\phi }  \label{RDef}
\end{equation}
is the operator of rotation in the spin space.

The general form of $\hat{H}_{A}$ is
\begin{equation}\label{ham-A}
\hat{H}_A = \hat{H}_{\parallel} + \hat{H}_{\perp}\,,
\end{equation}
where $\hat{H}_{\parallel}$ commutes with $S_z$ and
$\hat{H}_{\perp}$ is a perturbation that does not commute with
$S_z$. The existence of the magnetic anisotropy axis $Z$ means
that the $| \pm S\rangle$ eigenstates of $S_z$ are degenerate
ground states of $\hat{H}_{\parallel}$. Operator $\hat{H}_{\perp}$
slightly perturbs the $| \pm S\rangle$ states, adding to them
small contributions of other $|m_S\rangle$ states. We shall call
these degenerate normalized perturbed states $|\psi_{\pm
S}\rangle$. Physically they describe the magnetic moment of the
molecule looking in one of the two directions along the anisotropy
axis. Full perturbation theory with account of the degeneracy of
$\hat{H}_{A}$ provides quantum tunneling between the $|\psi_{\pm
S}\rangle$ states. The ground state and the first excited state
become
\begin{equation}\label{pm}
\Psi_{\pm} = \frac{1}{\sqrt{2}}\left(|\psi_{S}\rangle \pm
|\psi_{-S}\rangle\right)\,.
\end{equation}
They satisfy
\begin{equation}\label{Epm}
\hat{H}_A\Psi_{\pm} = E_{\pm}\Psi_{\pm}
\end{equation}
with $E_- - E_+ \equiv \Delta$ being the tunnel splitting.
Expressing $|\psi_{\pm S}\rangle$ via $\Psi_{\pm}$ according to
Eq.\ (\ref{pm}), it is easy to see from Eq.\ (\ref{Epm}) that
\begin{equation}\label{ME-A}
\langle \psi_{\pm S}|\hat{H}_A|\psi_{\pm S}\rangle =  0, \quad
\langle \psi_{-S}| \hat{H}_A|\psi_{S}\rangle = - {\Delta}/{2}\,.
\end{equation}
This gives
\begin{equation}\label{ME-S}
\langle \psi_{\pm S}|\hat{H}_S|\psi_{\pm S}\rangle =  0, \quad
\langle \psi_{\mp S}| \hat{H}_S|\psi_{ \pm S}\rangle = -
\frac{\Delta}{2}e^{\pm 2iS\phi}\,.
\end{equation}
for the matrix elements of $\hat{H}_S$.

Since the low-energy spin states of the molecule are
superpositions of $|\psi_{\pm S}\rangle$, it is convenient to
describe such a two-state system by a pseudospin 1/2. Components
of the corresponding Pauli operator ${\bm \sigma}$ are
\begin{eqnarray}\label{Pauli}
\sigma_x & = & |\psi_{-S}\rangle\langle \psi_{S}| +
|\psi_{S}\rangle\langle \psi_{-S}|
\\ \nonumber \sigma_y & = &
i|\psi_{-S}\rangle\langle \psi_{S}| - i|\psi_{S}\rangle\langle \psi_{-S}| \\
\nonumber \sigma_z & = & |\psi_{S}\rangle\langle \psi_{S}| -
|\psi_{-S}\rangle\langle \psi_{-S}|\,.
\end{eqnarray}
The projection of any operator $\hat{H}$ onto $|\psi_{\pm
S}\rangle$ states is given by
\begin{equation}\label{projection}
\sum_{m,n = \pm S}\langle
\psi_m|\hat{H}|\psi_n\rangle|\psi_m\rangle\langle \psi_n|\,.
\end{equation}
Noticing that
\begin{equation}
S_z|\psi_{\pm S}\rangle \cong S_z|\pm S\rangle = \pm S |\psi_{\pm
S}\rangle\,,
\end{equation}
it is easy to project Hamiltonian (\ref{ham}) onto $\psi_{\pm S}$.
Simple calculation yields
\begin{equation}
\hat{H}_{\mathrm{eff}}= \hat{H}_{\mathrm{rot}} -\frac{1}{2}W\sigma
_{z}-\frac{1}{2}\Delta \left[
\cos \left( 2S\phi \right) \sigma _{x}+\sin \left( 2S\phi \right) \sigma _{y}%
\right]   \label{HeffFinal}
\end{equation}
or
\begin{equation}
\hat{H}_{\mathrm{eff}}=
\hat{H}_{\mathrm{rot}}-\frac{1}{2}\mathbf{H}_{\mathrm{eff}}\cdot
{\bm \sigma }\,,
\end{equation}
where
\begin{equation}
\mathbf{H}_{\mathrm{eff}}=W\mathbf{e}_{z}+\Delta \cos \left(
2S\phi \right) \mathbf{e}_{x}+\Delta \sin \left( 2S\phi \right)
\mathbf{e}_{y} \label{HfieldeffDef}
\end{equation}
and
\begin{equation}
W=2Sg\mu _{B}B_{z}\,.  \label{WDef}
\end{equation}

Note that for a non-rotating magnetic molecule the effective
Hamiltonian has the form
\begin{equation}
\hat{H}_{\mathrm{eff}}=-\frac{1}{2}W\sigma _{z}-\frac{1}{2}\Delta
\sigma _{x}\,.  \label{HeffDef}
\end{equation}
Its eigenvalues are
\begin{equation}
E_{\pm }=\pm \frac{1}{2}\sqrt{W^{2}+\Delta ^{2}}  \label{EpmW}
\end{equation}
with
\begin{equation}
E_{+}-E_{-}\equiv \hbar \omega _{0}\equiv \sqrt{W^{2}+\Delta
^{2}}\,. \label{hbaromegaDef}
\end{equation}

Using the Heisenberg equation of motion for an operator $A(t)$,
\begin{equation}
\dot{A}(t)=\frac{i}{\hbar }\left[
\hat{H}_{\mathrm{eff}},A(t)\right]\,,
\end{equation}
one obtains
\begin{eqnarray}\label{dot}
& & \hbar \dot{L}_{z}=-I_{z}\omega _{r}^{2}\phi -S\Delta
\left[\sin \left( 2S\phi \right) \sigma _{x}-\cos \left( 2S\phi
\right) \sigma _{y}\right]\,, \nonumber \\
& &  \dot{\phi}=\frac{\hbar L_{z}}{I_{z}}\,,\quad
\mathbf{\dot{\sigma}=}\frac{1}{\hbar }\left[ \mathbf{\sigma \times H}_{%
\mathrm{eff}}\right]\,.
\end{eqnarray}
Elimination of ${L}_{z}$ provides the following system of
equations:
\begin{eqnarray}
\hbar \dot{\sigma}_{x} &=&W\sigma _{y}-\Delta \sin \left( \varphi
\right)
\sigma _{z}  \label{sigdot1} \\
\hbar \dot{\sigma}_{y} &=&\Delta \cos \left( \varphi \right)
\sigma
_{z}-W\sigma _{x}  \label{sigdot2} \\
\hbar \dot{\sigma}_{z} &=&\Delta \sin \left( \varphi \right)
\sigma _{x}-\Delta \cos \left( \varphi \right) \sigma _{y}
\label{sigdot3}
\end{eqnarray}
and
\begin{equation}
\ddot{\varphi}+\omega _{r}^{2}\varphi =\omega _{c}^{2}\left[ \cos
\left( \varphi \right) \sigma _{y}-\sin \left( \varphi \right)
\sigma _{x}\right] \,. \label{phiddotEq}
\end{equation}
Here
\begin{equation}\label{varphi}
\varphi \equiv 2S\phi
\end{equation}
and
\begin{equation}
\omega _{c} \equiv \sqrt{2S^{2}\Delta /I_{z}}\,.
\label{omegacDef}
\end{equation}

Combining the above equations it is easy to see that they satisfy
\begin{equation}\label{conservation}
\frac{d}{dt}(\hbar L_z + \hbar S\sigma_z) = -I_z\omega_r^2\phi\,.
\end{equation}
The left-hand side of this equation is the time derivative of the
$Z$-component of the total angular momentum, $\hbar(L_z + S_z)$,
while the right-hand side is the elastic torque due to the
coupling to the leads (see Fig.\ \ref{geometry}). This torque acts
such as to return the molecule to its equilibrium position, $\phi
= 0$. In the absence of such a torque the total angular momentum
of the molecule alone would be conserved.

Equations (\ref{sigdot1}) - (\ref{phiddotEq}) are operator
equations. To obtain numerically tractable equations, these
operator equations should be averaged over quantum states of the
system. If one decouples the quantum averages as
\begin{equation}
\left\langle \sin \left( \varphi \right) \sigma _{z}\right\rangle
\Rightarrow \left\langle \sin \left( \varphi \right) \right\rangle
\left\langle \sigma _{z}\right\rangle\,,
\end{equation}
in the spirit of the mean-field approximation, one obtains
classical-like equations of the same structure as above. Without
such a decoupling the equations for the Heisenberg operators are
useless and one has to go back to the Schr\"{o}dinger equation for
the whole system consisting of coupled spin and mechanical
subsystems. While the decoupling cannot be justified in the
general case, we notice that the cross terms containing ${\bm
\sigma}$ and $\phi$ become small when $\langle\varphi\rangle$ is
small. Since $\varphi = 2S\phi$ this condition for large $S$ is
stronger than the condition of small oscillations, $|\phi| \ll 1$.
Nevertheless, even for $S = 10$, which is the case of a Mn$_{12}$
molecule, it is likely that rotations of the molecule in the
geometry depicted in Fig.\ \ref{geometry} will still satisfy the
condition $|\varphi| \ll 1$. In this case Eqs.\ (\ref{sigdot1}) -
(\ref{phiddotEq}) can be treated as classical.

The behavior of the system depends on the dimensionless
magneto-mechanical constant of a free molecule,
\begin{equation}
\delta \equiv \left( \frac{\hbar \omega _{c}}{\Delta}\right) ^{2}=\frac{%
2\hbar ^{2}S^{2}}{I_{z}\Delta }\,,  \label{alphaDef}
\end{equation}
and the resonance parameter,
\begin{equation}
r\equiv \frac{\hbar \omega _{r}}{\Delta }\,. \label{rParDef}
\end{equation}
For a linear field sweep, $W = vt$, another relevant dimensionless
parameter is
\begin{equation}
\epsilon \equiv \frac{\pi \Delta ^{2}}{2\hbar v}\,.
\label{epsilonDef}
\end{equation}
It  determines probability,
\begin{equation}
P=e^{-\epsilon}\,,  \label{PLZ}
\end{equation}
of staying in the initial $\psi_{-S}$ state in the standard
Landau-Zener problem.

Using the above parameters and dimensionless variables
\begin{equation}\label{dimensionless}
\tau \equiv \frac{t\Delta}{\hbar}\, \quad w=\frac{vt}{\Delta
}=\frac{\hbar v\tau }{\Delta ^{2}}
=\frac{\pi \tau }{%
{2\epsilon}}\,,
\end{equation}
Eqs.\ (\ref{sigdot1}) - (\ref{phiddotEq}) for $|\varphi| \ll 1$
can be re-written in the form
\begin{eqnarray}\label{sigma-eq}
\sigma _{x}^{\prime } &=&w\sigma _{y}- \varphi \sigma _{z}
\nonumber
\\
\sigma _{y}^{\prime } &=&\sigma _{z}-w
\sigma _{x} \nonumber \\
\sigma _{z}^{\prime } &=&\varphi  \sigma _{x}-\sigma _{y}
\end{eqnarray}
and
\begin{equation}\label{varphi-eq}
\varphi ^{\prime \prime }+\gamma\varphi^{\prime}+r^{2}\varphi
=\delta \left[\sigma _{y}- \varphi  \sigma _{x}\right] \,,
\end{equation}
where prime means $d/d\tau$. Notice that in the last equation we
introduced a term $\gamma\varphi^{\prime}$ that describes damping
of the mechanical oscillations of the molecule.

\begin{figure}[ht]
\begin{center}
\vspace{-0.8cm}\hspace{-0.5cm}
\includegraphics[width=69mm,angle=-90]{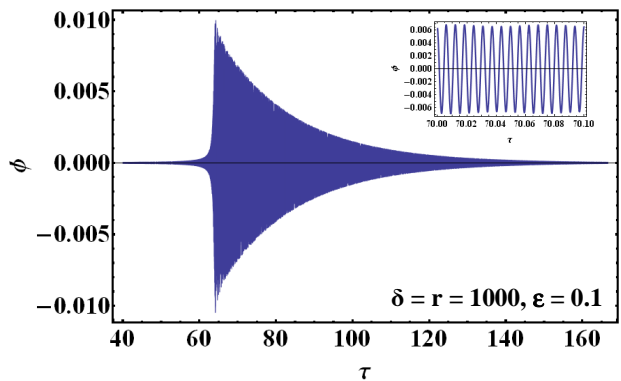}
\vspace{-1cm} \caption{Typical time dependence of the rotation
angle $\phi$. The inset shows the fine structure of the
oscillations.} \label{AnglePhiFastSweep}
\end{center}
\end{figure}
The above equations have been solved numerically using
Mathematica. With the Mn$_{12}$ molecule in mind, we compute the
time dependence of the angle of twist, $\phi={\varphi}/({2S})$,
for $S=10$, $\delta=10^{3}$ and $r=10^{3}$ (see discussion below).
For the illustration purpose, the value of the damping constant
$\gamma=0.1$ has been chosen to provide the finite duration of the
oscillations. Fig.\ \ref{AnglePhiFastSweep} shows the results for
a relatively fast field sweep corresponding to $\epsilon=0.1$. For
a slow sweep ($\epsilon \gg 1$) the amplitude of the oscillations
is significantly smaller.

Simple arguments allow one to understand the behavior shown in
Fig.\ \ref{AnglePhiFastSweep}. Oscillations of $\phi$ are excited
when the time-dependent distance between the spin levels given by
Eq.\ (\ref{hbaromegaDef}), coincides with the frequency $\omega_r$
of the mechanical oscillations of the molecule. For large
$\epsilon r$ this happens at $\tau =  2\epsilon r/\pi$ in
excellent agreement with Fig.\ \ref{AnglePhiFastSweep}.
Oscillations continue at the frequency $\omega_r$ until they are
completely damped due to the finite $\gamma$. To obtain the
dependence of the initial amplitude of oscillations on the
parameters the approximate solution of Eqs.\ (\ref{sigma-eq}) for
$\sigma_y(\tau)$ at small $\varphi$, $\epsilon < 1$ and $\tau
> 1$ can be used \cite{CGJ}:
\begin{equation}\label{sigma-y}
\sigma_y(\tau) =
-2\sqrt{e^{-\epsilon}(1-e^{-\epsilon})}\sin{\left(\frac{\pi\tau^2}{4\epsilon}\right)}\,.
\end{equation}
This gives an approximate solution of Eq.\ (\ref{varphi-eq}) at
$\gamma = 0$:
\begin{equation}\label{phi-int}
\varphi(\tau) = \frac{\delta}{r}{\rm
Im}\left[e^{ir\tau}\int^{\tau}_{0}d\tau'
e^{-ir\tau'}\sigma_y(\tau')\right]\,.
\end{equation}
The result of the integration can be expressed in terms of
trigonometric and error functions. While the formula is rather
cumbersome it provides a simple dependence of the final amplitude
of the undamped oscillations on the parameters:
\begin{equation}\label{amplitude}
\phi_{max} = \frac{2\hbar S}{I_z\omega_r}\sqrt{\epsilon
e^{-\epsilon}\left(1-e^{-\epsilon}\right)}\,.
\end{equation}
It reaches maximum at $\epsilon = 1.45$. This expression is in
excellent agreement with numerical results. It can be used as long
as $2S\phi_{max} \ll 1$.

Most of the existing experiments on transport through individual
magnetic molecules have been done with Mn$_{12}$ acetate. The
moment of inertia of the Mn$_{12}$ molecule is in the ballpark of
$10^{-34}$g$\cdot$cm$^2$. The tunnel splitting $\Delta$ depends
strongly on the transverse anisotropy, $\hat{H}_{\perp}$ in Eq.\
(\ref{ham-A}). For the molecule bridged between two leads it may
differ drastically from that in a Mn$_{12}$ crystal. With $I_z
\sim 10^{-34}$g$\cdot$cm$^2$ our choice of $\delta \sim 10^3$ in
Fig.\ (\ref{AnglePhiFastSweep}) corresponds to $\Delta/\hbar \sim
10^6$s$^{-1}$. The resonance frequency of the rotational
oscillations of the molecule must be much higher. The choice of $r
= 10^3$ in Fig.\ (\ref{AnglePhiFastSweep}) corresponds to
$\omega_r = 10^9$s$^{-1}$. The mechanical oscillations of the
molecule begin at $t = 2\epsilon\hbar^2\omega_r/(\pi \Delta^2)$
after the spin states $\psi_{-S}$ and $\psi_{S}$ cross due to the
field sweep. For the choice of parameters used in Fig.\
(\ref{AnglePhiFastSweep}) this time is of order $10^{-4}$s. Spin
oscillations given by Eq.\ (\ref{sigma-y}) must dissipate on a
longer time scale for the mechanical oscillations to be
observable. In a Mn$_{12}$ crystal, dissipation is dominated by
phonons. For a setup shown in Fig.\ \ref{geometry} phonon
processes must be suppressed, making spin relaxation times of
order $10^{-4}$s quite reasonable. For $\Delta/\hbar \sim
10^6$s$^{-1}$, the fast field sweep, $\epsilon <  1$, corresponds
to a few kOe per second, which is also reasonable. Magnetic
molecules other than Mn$_{12}$ acetate may prove to be even better
candidates for such an experiment.

If a tunneling current flows through the setup depicted in Fig.\
\ref{geometry}, the amplitude of the current should be affected by
the mechanical oscillations of the molecule.
Same as in tunneling microscopy, this effect should be detectable
due to high sensitivity of the electron tunneling rate to the
orientation of the molecule \cite{Voss}. Following the field
sweep, the current should acquire an oscillating component similar
to that shown in Fig.\ \ref{AnglePhiFastSweep}. Such an experiment
is practicable and it would be of great interest as it would prob
quantum dynamics of spin in individual magnetic molecules.

This work has been supported by the U.S. National Science
Foundation through Grant No. DMR-0703639.

\end{document}